\documentclass[twocolumn,aps,showpacs,prb,tightenlines,amsmath,amssymb,superscriptaddress]{revtex4}
\usepackage{graphicx}
\usepackage{amssymb}
\usepackage{dcolumn}
\usepackage{amsmath}
\usepackage{bm}

\usepackage{colordvi}

\begin{document}

\title{Density dependence of spin relaxation in GaAs quantum well at room temperature}
\author{L. H. Teng}
\affiliation{State Key Laboratory of Optoelectronic Materials
and Technologies, Department of Physics,
Zhongshan University, Guangzhou, Guangdong 510275, China}
\author{P. Zhang}
\affiliation{Hefei National Laboratory for Physical Sciences at
Microscale,
University of Science and Technology of China, Hefei,
Anhui, 230026, China}
\author{T. S. Lai}
\email{stslts@mail.sysu.edu.cn.}
\affiliation{State Key Laboratory of Optoelectronic Materials
 and Technologies, Department of Physics,
Zhongshan University, Guangzhou, Guangdong 510275, China}
\author{M.\ W.\ Wu}
\email{mwwu@ustc.edu.cn.}
\affiliation{Hefei National Laboratory for Physical Sciences at
Microscale,
University of Science and Technology of China, Hefei,
Anhui, 230026, China}
\affiliation{Department of Physics,
University of Science and Technology of China, Hefei,
Anhui, 230026, China}
\date{\today}

\begin{abstract}
Carrier density dependence of electron spin relaxation in an intrinsic GaAs
quantum well is investigated at room temperature using time-resolved
circularly polarized pump-probe spectroscopy. It is revealed that the
spin relaxation time first increases with density in the
relatively low density regime where the linear D'yakonov-Perel'
spin-orbit coupling terms are dominant, and then tends to decrease
when the density is large and the cubic D'yakonov-Perel'
spin-orbit coupling terms become important. These features are in good agreement
with theoretical predictions on density dependence of spin relaxation by L\"u {\em et al.} [Phys. Rev. B
{\bf 73}, 125314 (2006)].  A fully microscopic calculation
based on  numerically solving the kinetic spin Bloch equations
with both the D'yakonov-Perel' and the Bir-Aronov-Pikus mechanisms
included,  reproduces the
density dependence of spin relaxation very well.

\end{abstract}
\pacs{72.25.Rb, 78.67.De, 71.10.-w}

\maketitle
Spintronics is an intriguing and growing field which aims to incorporate the
spin degree of freedom into the traditional
electronics.\cite{zutic,fabian}
In this field, the spin relaxation time is one of the most important
basic  quantities, especially for the design of
spin-based devices. Typically, for example, ultrafast spin relaxation
process is needed for spin-dependent optical switch,\cite{nishikawa,hyland,hall}
while long enough spin lifetime is required when dealing with quantum information
storage as well as spin transport.\cite{kikkawa,kane,gershenfeld}
Investigations on spin relaxation in various materials and
structures have been carried out both experimentally and theoretically,
revealing that spin relaxation can be
affected/manipulated by various factors, such as
temperature,\cite{wu0,adachi,paillard,lv,zhou1,liu,holleitner,song} initial spin
polarization,\cite{stich} carrier and/or impurity
density,\cite{sandhu,zhou1,adachi,lombez,lv,liu,kikkawa,kikkawa1,wu0,song,dzhioev} magnetic
field,\cite{paillard,stich,holleitner,wu0} drift electric
field,\cite{barry,weng2} and so on.  Among these factors, the carrier density is one of the
most basic quantities which can be
easily controlled by gate voltage and/or optical excitation power.
Thus investigation on density dependence of
spin relaxation is necessary.

It is well known that for zinc-blende semiconductors such as GaAs,
the D'yakonov-Perel' (DP) mechanism is the leading mechanism of spin
relaxation.\cite{dyakonov} A femtosecond time-resolved Faraday
rotation measurement of $n$-type bulk GaAs has shown that the spin
relaxation time decreases with carrier density,\cite{kikkawa1} and the
theoretical calculation\cite{wu0} is consistent with this result. Note
that the spin-orbit coupling  terms (DP terms)
depend on momentum  cubicly in bulk
GaAs.\cite{dresselhaus,wu0} However, due to the confinement along the growth
direction, the DP terms in quantum wells include both
linear and cubic terms. For example, in (001) symmetric
GaAs quantum wells with small well width (so that only the
lowest subband is involved),
the DP terms come from the Dresselhaus terms:\cite{dresselhaus}
\begin{equation}
{\bf\Omega}({\bf
  k})=\gamma(k_x(k_y^2-\langle k_z^2\rangle), k_y(\langle
k_z^2\rangle-k_x^2), 0),
\end{equation}
 in which $\langle k_z^2\rangle$ stands for the
average of the operator $-(\partial/\partial z)^2$ over the electron
state of the lowest subband and $\gamma$ is the spin
splitting parameter. The relative importance of the linear and cubic DP terms
depends on the well width,\cite{weng2} temperature\cite{weng2} and
carrier density.\cite{lv}
The different momentum dependences of the DP terms lead to
complicated temperature and/or density dependences of spin
  relaxation in
GaAs quantum wells.
It is predicted from a fully microscopic kinetic spin Bloch equation (KSBE)
investigation\cite{zhou1} that in the strong scattering limit,   the
spin relaxatin time increases with carrier density
when the linear DP terms are dominant and decreases with density when
the cubic terms are important.\cite{lv} The
underlying physics is associated with the competition between the two
effects in the DP mechanism---the
inhomogeneous broadening\cite{weng2,lv} and the counter effect of
scattering on the broadening.\cite{weng2,lv}
Density and/or temperature can
affect this competition with the relative importance of each competing effect
depending on which part of the DP terms are dominant. In fact, a non-monotonic dependence of spin
relaxation on temperature in GaAs quantum wells was theoretically
predicted in Ref.~\onlinecite{weng2}, and was verified experimentally
recently.\cite{stich,holleitner} However, the predicted density dependence of
spin relaxation\cite{lv} has not yet been verified
experimentally. This work is to investigate the density dependence of
spin relaxation in (001) GaAs quantum wells at room
temperature. Unlike  some earlier studies on two
 dimensional GaAs with low carrier density and at low temperature 
where the excitonic effects dominate,\cite{bar} the present
investigation is in the regime of  the electron-hole plasma.

The two dimensional sample consists of 11 periods of 10 nm thick
GaAs quantum wells separated by 6 nm Al$_{0.3}$Ga$_{0.7}$As barriers,
grown on semi-insulating GaAs substrate by molecular beam epitaxy
along the (001) direction ($z$-axis). The substrate is removed by polishing
first and then selective chemical etching for transmission measurements. The substrate-free
 GaAs/AlGaAs films are mounted on a piece of sapphire window. The widely
used time-resolved circularly polarized pump-probe
spectroscopy\cite{tackeuchi,miller,tackeuchi1,hilton} is adopted to
realize spin pumping and spin-relaxation measurements.
 The femtosecond laser pulses generated
from a Ti:sapphire laser oscillator have a duration of 100\ fs, a half spectrum
width at half maximum of 6.7\ meV, and a repetition rate of 82\ MHz. By
passing through a standard time-resolved pump-probe setup, the pulses are split
into pump and probe ones with intensity ratio of 3:1.
The pump and probe beams are incident nearly normally to the sample
surface and focused by a lens of 50 mm focal length on
the sample to a spot size of about 30 $\mu$m in diameter. Two commercially available achromatic quarter-wave
 plates are inserted into the pump and probe beams, respectively, to
generate co-helicity or cross-helicity circularly polarized pump and
probe pulses. The differential transmission change of the probe is detected
 by a photodiode and measured by a lock-in amplifier which is referenced at
the modulation frequency of an optical chopper that modulates the pump beam. The central
wavelength of the pulses is tuned to 830\ nm to excite the heavy-hole
transition alone, and thus an initial degree of spin polarization of
nearly 100\ \% may be obtained.\cite{pfalz} In this experiment,
right circularly polarized
($\sigma^+$) pump pulses create spin-down
polarized electrons, while time-delayed $\sigma^+$ [left
circularly polarized ($\sigma^-$)] probe pulses measure the number
of spin-down (-up) electrons. Recombination of the
photo-excited carriers can be detected by using linearly polarized
light. In addition, a tunable optical attenuator
is used to control input laser pulse energy so that the electron density $N$
can change in a range of $0.3\times10^{11}$\ cm$^{-2}$ to $4\times10^{11}$\ cm$^{-2}$.
The excitation density is calculated by an usual formula, $(1-R)E\alpha/(h\nu S)$, with
 $R$ and $\alpha$ being the reflectivity and absorption coefficient of the sample, respectively. Here
$E$ is the pumping energy per pulse, $h\nu$ is the photon energy, and $S$ is the area of pump spot.

\begin{figure}[ht]
    {\includegraphics[width=9cm]{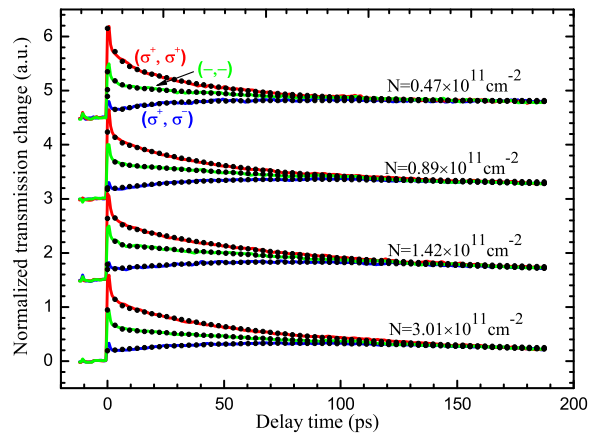}}
    \caption{(Color online) Normalized transmission change of probe
        beams under four typical carrier densities. The color solid
        curves are experimental data, while the black dots are the
        fittings with an elliptically
    polarized pump-probe absorption model. Curves labeled with
 ($\sigma^+$, $\sigma^+$) are
     taken from co-helicity pump-probe beams, with
  ($\sigma^+$, $\sigma^-$) are
      from the cross-helicity pump-probe beams and
with ($-$, $-$) are from collinearly polarized pump-probe beams.
Different $y$-axis
   offsets are added to each set of curves for clarity.}
  \label{fig1}
\end{figure}

\begin{figure}[ht]
    {\includegraphics[width=9cm]{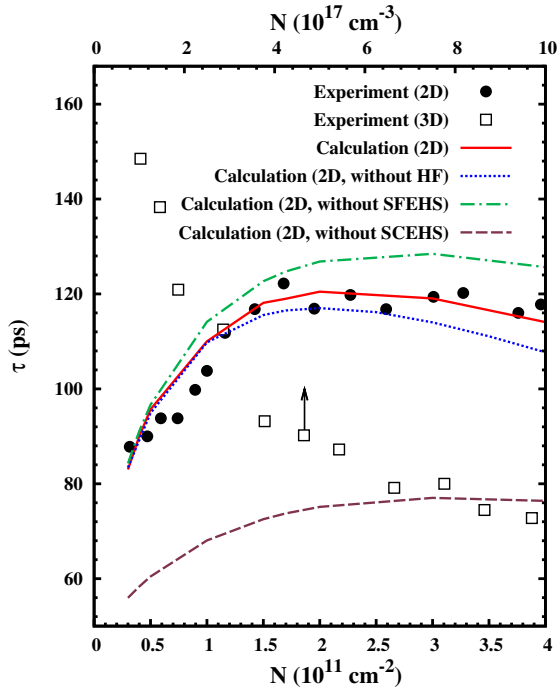}}
    \caption{(Color online) Carrier density $N$ dependence of electron
        spin relaxation time $\tau$. Dots: experimental data in
quantum well (2D); open squares: experimental data in
bulk material (3D). Solid curve: full theoretical
calculation; chain curve: theoretical calculation without
the SFEHS; dashed curve: theoretical calculation without
the SCEHS; dotted
 curve: theoretical calculation without the Coulomb HF term. Note the scale
of the bulk data is on the top frame of the figure.}
  \label{fig2}
\end{figure}

The main results are shown in Figs.\ \ref{fig1} and
\ref{fig2}. Figure~\ref{fig1} indicates the normalized time-delayed
scanning transmission change profiles of the probe beams for four
different excited carrier densities of
$0.47\times10^{11}$~cm$^{-2}$, $0.89\times10^{11}$~cm$^{-2}$,
$1.42\times10^{11}$~cm$^{-2}$

and $3.01\times10^{11}$~cm$^{-2}$, respectively. The
profiles labeled as $(\sigma^+,\sigma^+)$ and $(\sigma^+,
\sigma^-)$ are taken from co-helicity and cross-helicity circularly
polarized pump and probe beams respectively, whereas the ones labeled by
$(-,-)$ are the collinear polarization pump-probe traces which
describe the recombination processes of the photoexcited carrier
population. Initially, the $(\sigma^+,\sigma^+)$ profile is
stronger than the $(-,-)$ one, while the $(\sigma^+,\sigma^-)$
profile is weaker than  the $(-,-)$ one.  But finally both $(\sigma^+,
\sigma^+)$ and $(\sigma^+,\sigma^-)$ profiles tend to coincide
with the $(-,-)$ profile. This just shows the relaxation of spin
polarization between $|1/2\rangle$ and $|-1/2\rangle$ spin states of
electrons in conduction band.\cite{tackeuchi,lai} The $(\sigma^+,
\sigma^+)$ profile reflects the decay of majority spin population
directly photocreated by $\sigma^+$ pump pulses, while $(\sigma^+,
\sigma^-)$ profile rises initially toward the $(-,-)$ profile,
which reflects population increase in minority spin state induced by
spin flip from the majority spin state. An elliptically polarized
pump-probe spectroscopic model described in Ref.~\onlinecite{lai} is used to
fit the time-delayed experimental profile to retrieve spin relaxation
time $\tau$. The hole spin relaxation is irrelevant  because it is well known that
hole spin relaxation is very fast (in sub-picosecond time
scale).\cite{hilton} The results are shown
in Fig.\ \ref{fig2}, with dots corresponding to the results
in the quantum well. In addition, a similar experiment is performed on
bulk GaAs for comparison and the open squares in Fig.\ \ref{fig2} are
the results. It is found that with
the increase of carrier density, the spin relaxation time in
bulk material decreases monotonically with
carrier density, coinciding with the previous
reports in $n$-type bulk GaAs.\cite{kikkawa1,wu0}
However, in quantum wells the spin relaxation time first increases in
the low density regime and then tends to decrease  after
reaching a maximum of about 120\ ps at the density of
$1.7\times10^{11}$\ cm$^{-2}$.

In order to gain a deep insight into the experimental results
of the quantum well, we performed
 a fully microscopic KSBE calculation,\cite{zhou1} which takes account of all
relevant spin relaxation mechanisms [including both the DP
and Bir-Aronov-Pikus\cite{bir,zhou1} (BAP)
mechanisms].
The KSBEs constructed by the nonequilibrium Green function method
read\cite{zhou1}
\begin{equation}
\dot{\rho}_{\bf k}=\left.\dot{\rho}_{\bf k}\right|_{coh}
+\left.\dot{\rho}_{\bf
    k}\right|_{scat},
\end{equation}
in which $\rho_{\bf k}$  represent the
density matrices of electrons with momentum ${\bf k}$.
 $\dot{\rho}_{\bf k}|_{coh}$ are the coherent terms describing the
coherent spin precession due to the effective magnetic
fields from the DP term and
the Hartree-Fock (HF) Coulomb interaction and $\dot{\rho}_{\bf k}|_{scat}$
stand for the scattering terms.
In our calculation, all the relevant scatterings, such as
the electron--longitudinal optical phonon scattering, electron-electron
Coulomb scattering and electron-hole Coulomb scattering,
 are explicitly included.
The electron-hole Coulomb scattering
is further composed of both the spin-flip electron-hole scattering (SFEHS) and the
spin-conserving electron-hole scattering (SCEHS), with the former
leading to the spin relaxation due to the BAP mechanism. Expressions of the
coherent and scattering terms are given in
detail in Ref.~\onlinecite{zhou1}.

By solving the KSBEs, we obtain the spin relaxation time
as a function of photoexcited carrier
density. In the calculation, the spin splitting parameter $\gamma$
(as a fitting parameter) is chosen to be
21~meV$\cdot$nm$^3$.\cite{fabian} The initial electron
spin polarization is set to be 100~\%
following the experiment and the temperature is 300~K.
The solid curve in Fig.~\ref{fig2} is
from the full calculation which reproduces the experimental results very
well. The obtained results can be understood from the joint effect
 of the following
two competing effects: (i) With the increase of carrier density,
the spin conserving scattering
is strengthened. This tends to suppress the inhomogeneous
broadening from the momentum
dependence of the effective magnetic field (the DP terms)
by driving carriers to more homogeneous states in
  momentum space, and thus weakens the spin
relaxation in the strong scattering limit.\cite{weng2,lv,zhou1}
(ii) Both the inhomogeneous broadening and the SFEHS increase with the density. This leads to the
increase of spin relaxation. As pointed out by one of the authors in
Refs.~\onlinecite{weng2} and \onlinecite{lv}, when the linear $k$-dependence of the
DP term is dominant, the temperature and/or density dependence
of Effect~(i) is stronger than  that of Effect~(ii),
consequently the spin relaxation time increases with temperature and/or carrier
density. However, when the cubic term becomes dominant,
the increase of inhomogeneous broadening [Effect~(ii)] with the
temperature and/or density overcomes  Effect~(i),
consequently, the spin relaxation time decreases with temperature and/or
carrier density. This is exactly what happens in
Fig.~\ref{fig2}: When $N<1.5\times 10^{11}$~cm$^{-2}$, the
linear DP term dominates and the spin
relaxation time increases with carrier density. When
the density goes higher, the spin relaxation time tends to
decrease. However, this decrease is moderate
as the contribution from the cubic DP term has not yet become dominant but is
comparable with the linear term at the present photoexcited density. This can be further seen from the chain curve
where the SFEHS, i.e., the BAP term, is turned off. The decrease
becomes even milder. Moreover, by comparing the solid curve and the
chain curve, one finds the spin relaxation from the BAP mechanism
becomes stronger with the increase of photoexcited carrier (especially
hole) density. However, the spin relaxation is still dominated by the DP
mechanism, as addressed very recently by Zhou and Wu in Ref.~\onlinecite{zhou1}.
We further show that for  intrinsic sample, due to the
same electron and hole densities,
the SCEHS makes marked contribution to the
spin relaxation due to the DP mechanism. This can be seen
from the dashed curve where the SCEHS is turned off.
One obtains much shorter spin relaxation time as the counter effect of the
scattering to the inhomogeneous broadening is markedly weakened by neglecting
the SCEHS.

Finally we address the issue of the initial spin polarization.
Due to the intrinsic two-dimensional sample, the initial  spin
polarization is 100~\% by the circular polarized laser excitation.
Unlike the low temperature case where an effective magnetic field
in Faraday configuration is induced by the Coulomb HF interaction
and the spin relaxation is markedly reduced,\cite{stich}
here the effective magnetic field is very small due to the high
temperature (so the electron distribution functions are much smaller than 1
for most momentums).
Therefore, the effect of the HF term to the spin relaxation is marginal,
as shown by the dotted curve in Fig.~\ref{fig2} where the HF term is
removed. It is further seen from the figure that with the increase
of the density, the contribution of the HF term becomes noticeable and
the spin relaxation is suppressed, in agreement with the
previous theoretical predictions and experimental observations.\cite{stich}

In summary, the carrier density dependence of electron spin relaxation
in intrinsic (001) GaAs quantum wells  is investigated by
a femtosecond pump-probe experiment at room temperature.
The spin relaxation time shows an
obvious increase with density in the relatively low density
regime and then a mild decrease  when the density is larger,
which is in good agreement with the theoretical predictions.\cite{lv}
 Further calculation with the fully microscopic KSBE approach reproduced the
experimental results very well. It is understood that in the
strong scattering limit,
when the carrier density is low and thus the linear DP term is dominant,
the spin relaxation time increases with density
due to the increasing counter effect of the scattering on
the inhomogeneous broadening. However, when the
density is large enough and the cubic DP term becomes important, the spin
relaxation time tends to decrease with density, thanks to the
rapid increase of the inhomogeneous broadening and the enhanced effect from
the BAP mechanism.

This work was supported by the Natural Science Foundation of China
under Grant Nos.\ 10574120, 10725417, 60490295, and 60678009, the
National Basic Research Program of China under Grant No.\
2006CB922005 and the Knowledge Innovation Project of Chinese Academy
of Sciences.

\end{document}